\documentclass{sigplanconf}

\usepackage{amsmath}
\usepackage{colortbl}
\usepackage{multirow}
\usepackage{graphicx}

\begin{document}

\setlength{\pdfpageheight}{\paperheight}
\setlength{\pdfpagewidth}{\paperwidth}

\conferenceinfo{CONF 'yy}{Month d--d, 20yy, City, ST, Country} 
\copyrightyear{20yy} 
\copyrightdata{978-1-nnnn-nnnn-n/yy/mm} 
\doi{nnnnnnn.nnnnnnn}

% Uncomment one of the following two, if you are not going for the 
% traditional copyright transfer agreement.

%\exclusivelicense                % ACM gets exclusive license to publish, 
                                  % you retain copyright

%\permissiontopublish             % ACM gets nonexclusive license to publish
                                  % (paid open-access papers, 
                                  % short abstracts)

\titlebanner{banner above paper title}        % These are ignored unless
\preprintfooter{short description of paper}   % 'preprint' option specified.

\title{A Preliminary Field Study of Game Programming on Mobile Devices}
%\subtitle{Subtitle Text, if any}

\authorinfo{Eric Anderson}
			{Department of Computer Science\\
           North Carolina State University}
           {ehander3@ncsu.edu}
\authorinfo{Sihan Li\and Tao Xie}
			{Department of Computer Science\\
           University of Illinois at Urbana-Champaign}
           {\{sihanli2, taoxie\}@illinois.edu }

\maketitle

\begin{abstract}
TouchDevelop is a new programming environment that allows users to create applications on mobile devices. Applications created with TouchDevelop have continued to grow in popularity since TouchDevelop was first released to public in 2011. This paper presents a field study of 31,699 applications, focusing on different characteristics between 539 game scripts and all other non-game applications, as well as what make some game applications more popular than others to  users. The study provides a list of findings on characteristics of game scripts and also implications for improving end-user programming of game applications.
\end{abstract}

%\category{CR-number}{subcategory}{third-level}

% general terms are not compulsory anymore, 
% you may leave them out
\terms
Languages, Human Factors.


\begin{table*}
%\small
\centering
\caption{Findings and Implications on TouchDevelop Game Scripts}
\renewcommand{\arraystretch}{1.21}
\begin{tabular}{|c|c|}
\hline
\rowcolor[gray]{0.8}
\textbf{Comparison of game scripts and non-game scripts} & \textbf{Implications} \\
\hline
\multicolumn{1}{|p{8.5cm}|}{TouchDevelop game scripts are typically larger than non-game
scripts.} &
\multicolumn{1}{p{8.7cm}|}{It could be important to provide effective IDE support for users to create complex and long game scripts.}  \\ 
\hline
\multicolumn{1}{|p{8.5cm}|}{TouchDevelop game scripts tend to be more popular than non-game
scripts.} &
\multicolumn{1}{p{8.7cm}|}{
%A more convenient system for giving positive or negative reviews may increase user feedback. 
In order to attract more users, it would be advantageous for TouchDevelop to promote and market game scripts as a very important type of scripts.} \\
\hline
\multicolumn{1}{|p{8.5cm}|}{ TouchDevelop game scripts typically have more actions and events
than non-game scripts.} &
\multicolumn{1}{p{8.7cm}|}{ Games tend to have more features and are more complex than non-games. It could be important to provide effective IDE support to encourage creation of advanced scripts. } \\
\hline
\multicolumn{1}{|p{8.5cm}|}{Game scripts typically invoke more methods
than non-game scripts.} &
\multicolumn{1}{p{8.7cm}|}{ Game scripts tend to include more features than non-game scripts and may also be more complex than non-game scripts. } \\
\hline
\rowcolor[gray]{0.8}
\textbf{Comparison of popular and unpopular game scripts} & \textbf{Implications}\\
\hline
\multicolumn{1}{|p{8.5cm}|}{Popular TouchDevelop game scripts tend to be slightly longer than
less popular scripts.} &
\multicolumn{1}{p{8.7cm}|}{ Users prefer scripts longer games, as a longer length allows for more features to be built into a game. It would be advantageous for TouchDevelop to expand the IDE to encourage the creation of larger and more complex scripts.}\\
\hline
\multicolumn{1}{|p{8.5cm}|}{Popular TouchDevelop game scripts use platform features such as accelerometer, music, and sounds more frequently than less popular
game scripts. } &
\multicolumn{1}{p{8.7cm}|}{TouchDevelop users prefer feature-rich games with platform features such as accelerometer and music. It could be helpful to integrate tutorials into the TouchDevelop environment for teaching users to use music and sounds, accelerometer and other popular platform features.} \\
\hline
\multicolumn{1}{|p{8.5cm}|}{Popular TouchDevelop game scripts seem to have slightly more actions and events than less popular game scripts, although this difference may not be statistically significant. } &
\multicolumn{1}{p{8.7cm}|}{Popular game scripts and unpopular game scripts seem to have similar numbers of events and actions, indicating that a large number of actions and events may not be a contributing factor to the popularity of a TouchDevelop game script.} \\
\hline
\end{tabular}
\vspace{-2mm}
\label{tbl: findings}
\end{table*}

\section{Introduction}

In the past few years, the smartphone market has been growing rapidly. According to Blodget [1], 114 million out of the 310 million people in the United States used smartphones in July 2012, just over 10 years after the first smartphone was released in the United States. Svensson [2] shows that smartphone sales overtook the sales of regular mobile phones in early 2013, with smartphones now comprising 51.6 percent of the worldwide mobile-phone market. With the prevalence of smartphones, applications for these smartphones are quickly becoming popular and are currently in high demand. For example, according to a press release issued by Apple [3], their app store alone features almost half a million apps, and more than 15 billion apps had been released as of July 2011. With the rapid growth of the smartphone-app industry, that number is expected to be  much higher today.

The rise of the smartphone-app market comes along with the rise of the smartphone-gaming market. Smartphones feature thousands of games available for download, including popular games also released for traditional game consoles. In fact, one developer reported that the version of their game released for the iPad was sold faster than the version released for common consoles and computers [4].
	
Typically, most of applications, including games, are written in object-oriented programming languages such as C\#, Objective-C, and Java.
%as can be seen in an article published recently [5]. 
Users must circumvent various barriers faced when using these advanced programming languages for creating mobile applications, and also they need to install development environments and emulators  in order to properly create mobile application and port it over to a mobile environment. These barriers make the process of creating applications difficult, especially for new users. 

TouchDevelop [5], a new development environment created by Microsoft Research, seeks to lower these barriers. With TouchDevelop, users can create applications directly on their mobile device, using a relatively simple programming language. TouchDevelop allows users of all skill levels to create a range of applications (i.e., TouchDevelop scripts), including games, which are the focus of this study, productivity tools, customized gadgets, and so on. In addition, TouchDevelop also provides an online bazaar that allows the public to share their applications. Users can not only download and run these applications but also obtain the source code of these applications.
Currently, there are thousands of TouchDevelop scripts created and published by users. 
% Additionally, all TouchDevelop code is available to the public free of charge, such that data-mining using TouchDevelop is a simple process.

Given the popularity of TouchDevelop, there is already some exiting work that studies end-user programming on mobile devices. Li et al. [6] analyzed the applications created in TouchDevelop and the users in TouchDevelop in order to discover what makes TouchDevelop scripts unique and identify common features among scripts. Athreya et al.  [7] studied what kinds of scripts can be created with TouchDevelop and what are problems faced by users. However, these two studies focus on general applications, without delving into a specific type of application such as game applications. Since game applications are one of the most popular types of applications, it would be valuable to conduct in-depth studies on game applications. 
	
In this study, we analyze a specific type of applications created in TouchDevelop: games. In particular, we intend to investigate how game scripts are different than non-game scripts, as well as what characteristics are common among popular game scripts. Such findings can give insights on providing better tool support to TouchDevelop users. For example, if there is a certain feature prevalent in  most popular game scripts, it would be important to develop tutorials to teach new users how to implement this feature. In particular, we intend to answer the following questions:

\textbf{RQ1: What do TouchDevelop game scripts look like when compared to non-game scripts?} We intend to compare various metrics of games and non-games, such as lines of code (LOC), frequency of use, and usage of platform features.

\textbf{RQ2: How do games compare to non-games in terms of popularity?} We intend to see how frequently TouchDevelop users are using games vs. non-games, based on metrics such as the number of installations, number of runs, and number of positive reviews. Answers to this question can help highlight where the interest of TouchDevelop users lies and what types of scripts are in high demand.

\textbf{RQ3: What features do more popular games have while less popular games lack?} We intend to discover reasons why some games are more popular than others during a given time period , in order to provide insights on tool support for making learning recommendations based on important features found in popular game scripts.

Table I lists all the findings and implications from our study. We describe the details of our study in the rest of the paper. Our paper makes the following main contributions:
\begin{itemize}
\item We present the first in-depth study on game scripts created on mobile devices. We provide valuable implications for improving end-user programming of games.
\item Our study findings show that game scripts are more complex and longer than other non-game scripts. In addition, the complexity of an application and its use of platform features could be factors that contribute to the popularity of game scripts.
\end{itemize} 

The rest of the paper is organized as follows. Section 2 describes the methodology used to conduct our study; section 3 presents details of our findings and implications; section 4 presents related work and section 5 concludes the paper.

\section{Methodology}
%In this section, we present the details of our methodology to conduct the study. We first describe how do we obtain the subjects and then describes metrics are used in our study to get our
\subsection{Subjects}
The subjects in this study were all TouchDevelop scripts published before the time of our final experiment (early June, 2013). Overall, there were 31,699 scripts included in the study. For each script, we downloaded the source code, id of the script, and publication time of the script, as well as other information like the platform features used by the script and the number of positive reviews received by that script from other TouchDevelop users. Of the 31,699 scripts analyzed, 539 were game scripts, and these scripts were analyzed separately from the rest of the scripts, although we downloaded the same information from game and non-game scripts. All data were obtained from the TouchDevelop cloud [8].

\subsection{Metrics and Approaches}
To answer RQ1 to determine how TouchDevelop game scripts compare to TouchDevelop non-game scripts, we used a number of metrics, listed as follows: number of methods called, number of actions, number of events, number of lines of code, and the prevalence of certain platform features, such as sound and accelerometer. We calculated both mean and median values for each of these metrics, excluding the prevalence of certain platform features, and then compared the mean and median values of game scripts to the mean and median values of non-game scripts. For the prevalence of certain platform features, we calculated the percentage of scripts that contained that platform and then compared percentages of game scripts and non-game scripts. 

To answer RQ2 about the popularity of game scripts compared to the popularity of non-game scripts, we looked at the number of times a script was run, assuming that more popular scripts would be run many times. In order to account for the fact that older games would have more runs simply because they had been around for a long time, only scripts published before July 2012 were looked at. We then found the mean and median number of runs for the game scripts and the non-game scripts and compared the values.

To answer RQ3 about the important features of popular game scripts as opposed to unpopular games scripts, we looked at a number of metrics, including number of methods called, number of actions, number of events, length of scripts, and the prevalence of music and accelerometer. A script was defined as “popular” if its number of runs was in the 75th percentile or greater. We first determined which game scripts were “popular” and which game scripts were “unpopular” and then analyzed the metrics listed above in a similar fashion to the analysis of RQ1, by finding the mean and median values of the metrics and then comparing them to one another.

\begin{figure}
\centering
\includegraphics[width=0.45\textwidth]{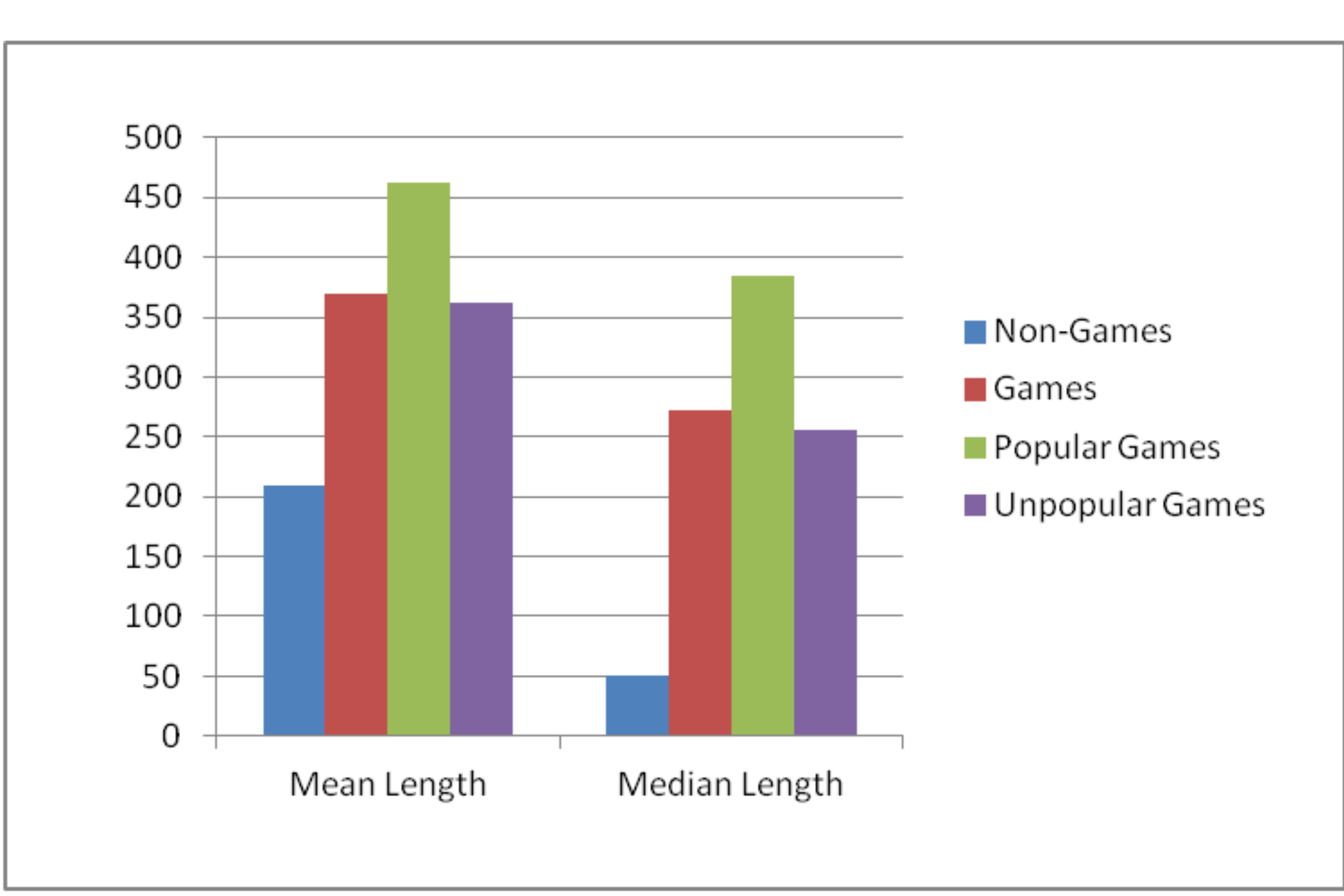}
\caption{Mean and median script length among various types of scripts.
}
\label{fig1}
\end{figure}

\section{Findings and Implications}
Figure~\ref{fig1} shows the various mean and median lengths of scripts among games and non-games. It can be seen that games are typically longer than non-games by almost 200 lines of code. Additionally, non-game scripts are close in length to the average length among all scripts, which is 213 lines of code, and game scripts are also much longer on average than scripts in general. This result is to be expected, as game scripts form a very small minority of scripts; only 1.7\% of scripts are game scripts. Game scripts may seem to be very small regardless. However, as most TouchDevelop scripts are extremely small due to being developed on mobile phones, game scripts are actually very long by comparison.

\textbf{Finding 1:} TouchDevelop game scripts are typically larger than non-game scripts. The average length of non-game scripts is 210 lines of code and the average length of game scripts is 369 lines of code. Additionally, the median size of non-games is 50 lines of code and the median size of games is 272 lines of code.

\textbf{Implication:} Games tend to be longer than their non-game counterparts. 
It could be important to provide effective IDE support for users to create complex and long game scripts.
%In order to allow users to create complex and longer game scripts, it is important to expand IDE support for TouchDevelop.

Even though they comprise such a small portion of all TouchDevelop scripts, game scripts are significantly more popular than non-game scripts. They have almost 10 times as many runs and installations on average in addition to about six times as many positive reviews. Additionally, it can be seen that scripts in general do not tend to receive many positive reviews, as non-games receive less than an average of half of a positive review, and games only receive an average of 1.8 positive reviews per script. This result would suggest that most TouchDevelop users do not frequently review the scripts that they run, and a change in how scripts are reviewed could benefit TouchDevelop.

\textbf{Finding 2:} TouchDevelop game scripts tend to be more popular among users than non-games. Non-game scripts have an average of 23.5 runs, 3.2 installations and 0.3 positive reviews, while game scripts have an average of 216.0 runs, 35.0 installations, and 1.8 positive reviews.

\textbf{Implication:} TouchDevelop users tend to be more interested in games and entertainment than  utility/productivity applications. Additionally, users seem to seldom review the scripts that they are using. A more convenient system for giving positive or negative reviews may increase user feedback. In order to attract more users, it would be advantageous for TouchDevelop to promote and market game scripts as an important type of scripts.

Our previous finding that game scripts are both longer than and more popular than non-game scripts on average might suggest a link between the popularity of a script and its length, and we also find it to be true among game scripts. Popular game scripts are found to have about 100 more lines of code on average than less popular game scripts, and the median among popular game scripts is also about 130 lines of code greater than the median among non-game scripts. 

\textbf{Finding 3:} Popular TouchDevelop game scripts tend to be slightly longer than less popular scripts. Popular scripts have a mean length of 462 lines of code and a median length of 384 lines of code, whereas unpopular scripts have a mean length of 363 lines of code and a median length of 255 lines of code.

\textbf{Implication:} Users prefer longer games, as a longer length allows for more features to be built into the game. It would be advantageous for TouchDevelop to expand the IDE to encourage the creation of larger and more complex scripts.

As would be expected, popular and better game scripts use platform features such as music and sounds and accelerometer more frequently than unpopular game scripts. The accelerometer especially seems to be fairly common among popular game scripts as it is featured in 37.1\% of those scripts. This result is to be expected, as the use of accelerometer takes advantage of the mobile gaming platform in a fairly unique way and would therefore be attractive to many users.

\textbf{Finding 4:} Popular TouchDevelop game scripts use platform features such as accelerometer and music and sounds more frequently than less popular game scripts. 37.1\% of popular game scripts feature accelerometer compared to 24.2\% of unpopular game scripts. Additionally, 20.0\% of popular game scripts feature music and sounds compared to 14.5\% of unpopular game scripts.

\textbf{Implication:} TouchDevelop users prefer feature-rich games with platform features such as accelerometer and music. To help users create scripts that contain these features, it could be helpful to integrate tutorials into the TouchDevelop website for teaching users to use music and sounds, accelerometer and other popular platform features.

\begin{figure}
\centering
\includegraphics[width=0.45\textwidth]{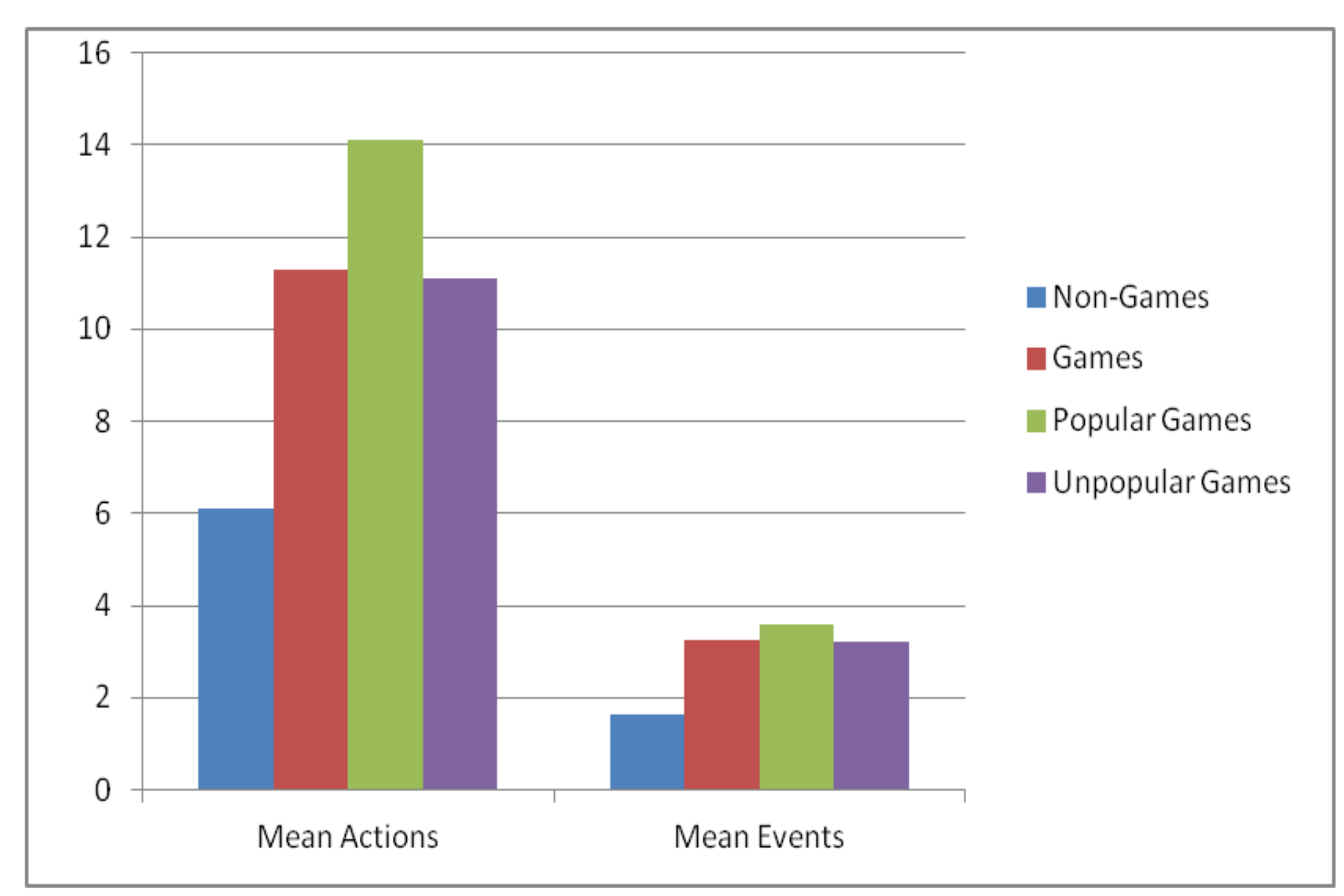}
\caption{Mean number of actions and events per different types of scripts.
}
\label{fig2}
\end{figure}

The mean number of actions and events can be seen in Figure~\ref{fig2}. Similar to Finding 3 on the large number of lines of code, a large number of actions and events seems to be common among games as opposed to non-games, as games have on average almost twice as many actions and events as non-games.

%\begin{table}
%\centering
%\caption{Mean method calls and method calls per line in game scripts and non-game scripts}
%\begin{tabular}{|c|r|r|}
%\hline 
%\textbf{Type} & \textbf{Mean MC. per script} & \textbf{Mean MC. per LOC} \\
%\hline
%Game & 347.18 & 0.94 \\
%\hline
%Non-game & 165.38 & 0.79 \\
%\hline
%\end{tabular}
%\label{tbl: tbl2} 
%\end{table}

\textbf{Finding 5:} TouchDevelop game scripts typically have more actions and events than non-game scripts. Non-game scripts on average have 6.1 actions and 1.6 events while game scripts on average have 11.3 actions and 3.2 events, which is almost twice as many actions and events per script.

\textbf{Implication:} TouchDevelop games tend to have more features and are more complex than non-games, possibly indicating that creators of complicated scripts tend to make games rather than other scripts. As stated previously, expanding the IDE could encourage creation of advanced scripts.

Because game scripts tend to feature more actions and events than non-game scripts, it would be expected that popular game scripts feature more actions and events than unpopular game scripts. However, it can be seen in Figure~\ref{fig2} that this expectation is not true. Popular and unpopular game scripts have very similar numbers of actions and events, suggesting that a large number of actions and events is common in all game scripts, not just in popular game scripts.

\textbf{Finding 6:} TouchDevelop game scripts typically have more methods called than non-game scripts. In average, a game script calls 347 methods whereas a non-game script calls 165 methods. Additionally, in average, a game script calls 0.9 methods per line of code compared to 0.8 methods per line of code in a non-game script. This result shows that games tend to have more methods called in total and more methods called per line.

\textbf{Implication:} Game scripts tend to include more features than non-game scripts and are also more complex than non-game scripts. In order to encourage users to create more complex scripts and games, it would be advantageous
for TouchDevelop to promote and market game scripts as an important type of scripts. Meanwhile, a good IDE could also help users create more complex scripts and games.

From our study results, game scripts seem to feature more method calls than non-game scripts, both as a whole and per line of code. This result further supports the conclusion that game scripts tend to contain more features and are more complex than non-game scripts.

\textbf{Finding 7:} Popular TouchDevelop game scripts have slightly more actions and events than less popular game scripts, although this difference may not be statistically significant. Popular scripts have an average number of 14.1 actions compared to an average of 11.1 actions per less popular script, but both popular and unpopular scripts have a median of 8.0 actions per scripts. Popular scripts also have a mean of 3.6 events per script, compared to an average of 3.2 events per non-popular script, although this difference seems negligible. Additionally, the median number of events for popular game scripts is 2 compared to a median number of 1 for unpopular game scripts.

\textbf{Implication:} Popular game scripts and unpopular game scripts seem to have similar numbers of events and actions, indicating that a large number of actions and events may not be a contributing factor to the popularity of a TouchDevelop game script.

\section{Related Work}
The most related work is a previous study on all TouchDevelop scripts conducted by Li et al. [6]. That study, similar to ours, focuses primarily on the structural features of scripts. However, that study has a more general scope, whereas this study focuses primarily on game scripts. Additionally, that study also investigates the users that create the scripts, whereas our study does not. Athreya et al. [7] presents a comprehensive study of TouchDevelop and finds that a plurality of scripts are related to entertainment, though they also find that a large number of scripts that do not have any functionality.

There is some other work that includes an introduction and summary of TouchDevelop [5] as well as a report on the accessibility of TouchDevelop for high school and middle school students [9], [10]. The report finds great success with teaching high school and middle school students using TouchDevelop. 

\section{Conclusion}
In this paper we have presented a study of game scripts for mobile devices using TouchDevelop. We studied 31,699 non-game scripts and 539 game scripts and compared them to each other to help determine what distinguishes game scripts from non-game scripts as well as what makes game scripts popular. We found that 1) TouchDevelop game scripts are longer than non-game scripts; 2) TouchDevelop game scripts are more complex and contain more actions and methods called than non-game scripts; 3) Popular TouchDevelop game scripts are typically longer and more complex than less popular game scripts. Based on these findings, we recommended an expansion of the TouchDevelop IDE and an expansion of TouchDevelop tutorials, to better teach and encourage the creation of complex scripts by users.

\section*{Acknowledgments}
This work is supported in part by NSF grants CCF-0845272, CCF-0915400, CNS-0958235, CNS-1160603, CCF-1349666, CNS-1318419, a Microsoft Research Award, and NSF of China No. 61228203. We thank Nikolai Tillmann for his help and feedback in conducting the study described in this paper.

%\appendix
%\section{Appendix Title}
%
%This is the text of the appendix, if you need one.

%\acks
%
%Acknowledgments, if needed.

% We recommend abbrvnat bibliography style.

\bibliographystyle{abbrvnat}

% The bibliography should be embedded for final submission.

\end{document}